\documentclass[reprint,aps,amsmath,amssymb,floatfix,screen]{revtex4-1}
\usepackage{eucal,graphicx,color,dsfont,mathrsfs,booktabs}
\usepackage[normalem]{ulem}
\usepackage{amsmath,amssymb,bm,mathrsfs}
\usepackage{srcltx}
\usepackage{dsfont}
\usepackage{epsfig}
\usepackage{slashed}
\usepackage{bbold}
\usepackage{psfrag}
\usepackage{xcolor}
\PassOptionsToPackage{caption=false}{subfig}
\usepackage{subfig}
\usepackage{xfrac}
\usepackage{multirow}
\usepackage{booktabs}
\usepackage[colorlinks=true,linkcolor=blue,citecolor=blue,urlcolor=violet]{hyperref}
\usepackage{mathtools}

\usepackage{soul}
\begin{document}

\title{Inferring the quantum density matrix with machine learning}

\newcommand{\be}{\begin{equation}}
\newcommand{\ee}{\end{equation}}
\newcommand{\bea}{\begin{eqnarray}}
\newcommand{\eea}{\end{eqnarray}}
\renewcommand{\r}{\rho}
\renewcommand{\l}{\lambda}
\newcommand{\ket}[1]{|#1\rangle}
\newcommand{\bra}[1]{\langle#1|}
\newcommand{\proj}[1]{\ket{#1}\bra{#1}}
\newcommand{\Tr}{{\textrm{Tr\,}}}
\newcommand{\KL}[2]{D_{\text{KL}}({#1}\parallel{#2})}
\newcommand{\QRE}[2]{S({#1}\!\parallel{\!#2})}
\newcommand{\xxx}[1]{{\color{red}\bf[#1]}}
\newcommand{\yyy}[1]{{\color{blue}\bf[#1]}}

\author{Kyle Cranmer}
\email{kyle.cranmer@nyu.edu}
\affiliation{
Center for Cosmology and Particle Physics,
Department of Physics, New York University, New York, NY 10003, USA.
} 

\author{Siavash Golkar}
\email{golkar@nyu.edu}
\affiliation{
Center for Cosmology and Particle Physics,
Department of Physics, New York University, New York, NY 10003, USA.
} 

\author{Duccio Pappadopulo}
\email{duccio.pappadopulo@gmail.com}
\affiliation{Bloomberg LP, New York, NY 10022, USA
} 


\begin{abstract}
We introduce two methods for estimating the density matrix for a quantum system: Quantum Maximum Likelihood and Quantum Variational Inference. In these methods, we construct a variational family to model the density matrix of a mixed quantum state. 
We also introduce  \emph{quantum flows}, the quantum analog of normalizing flows, which can be used to increase the expressivity of this variational family. The eigenstates and eigenvalues of interest are then derived by optimizing an appropriate loss function. 
The approach is qualitatively different than traditional lattice techniques that rely on the time dependence of correlation functions that summarize the lattice configurations. The resulting estimate of the density matrix can then be used to evaluate the expectation of an arbitrary operator, which opens the door to new possibilities. 
\end{abstract}

\maketitle

\section{Introduction}\label{sec:intro}

There is a nexus of concepts at the heart of a rich interplay between physics, statistics, machine learning, and information theory. Concepts such as entropy  that were key to the early work in thermodynamics are the bedrock of information theory. Similarly the Gibbs (or Boltzman) distribution, which characterize the distribution of states in thermal equilibrium, is at the heart of energy based models and Boltzman machines that were widely studied in machine learning~\cite{ackley1985learning,LeCun06atutorial}. Additionally, the study of complicated many-body systems gave rise to mean-field methods and renormalization group methods. In particular, the 
Gibbs-Bogoliubov-Feynman inequality, which provides a lower-bound on the intractable partition function of these complicated systems reappears in Bayesian statistics in the form of the evidence lower bound (ELBO) central to Variational Inference~\cite{peterson1989explorations,wainwright2008graphical}. 

Shortly after quantum mechanics was developed, many of these core concepts found their quantum analogs. The extension of classical Shannon entropy to quantum systems, known as the von Neumann entropy, requires a notion of mixed states. Mixed states combine the uniquely quantum mechanical concept of a coherent superpositions and the more classical notion of an incoherent superposition. Importantly, mixed states are needed to generalize the notion of a Gibbs distribution to a quantum system in thermal equilibrium.

For decades these concepts have been refined and extended forming an enormous body of work in the respective fields of study. Not surprisingly, as the fields have specialized, the language has diverged, and the corresponding jargon has become a barrier to cross-fertilization. 

With the rise of deep learning in the last few years,  there has been a surge of research connecting machine learning methods to problems in physics~\cite{Carleo:2019ptp}.  In particular, machine learning techniques have been used for variational optimization of ground state energy for quantum systems~\citep{carleo_solving_2017}. 
Additionally, there have been a number of important developments that extend statistical inference to domains where probabilistic modeling was previously inaccessible.
These techniques have recently been explored to solve statistical mechanics of classical systems~\cite{PhysRevLett.122.080602,2019arXiv190311048N}.
 In this work, we aim to connect recent developments in deep generative models~\cite{larochelle2011neural,papamakarios2017masked,2015arXiv150505770J, DBLP:journals/corr/DinhSB16}, unsupervised learning for implicit models~\cite{goodfellow2014generative}, and variational inference~\cite{ranganath2014black} to their quantum mechanical analogs. We hope this will bring a fresh lens to well studied problems and encourage new approaches leveraging the recent advancements in deep learning.

In this paper, we introduce two methods for estimating the density matrix for a quantum system: Quantum Maximum Likelihood (QML) and Quantum Variational Inference (QVI). In these methods, we model the density matrix for a mixed state with a variational family. 
We also introduce  \emph{quantum flows}, the quantum analog of normalizing flows~\cite{larochelle2011neural,papamakarios2017masked,2015arXiv150505770J, DBLP:journals/corr/DinhSB16}, which can be used to increase the expressivity of this variational family. The eigenstates and eigenvalues of interest are then derived by optimizing an appropriate loss function. 
The corresponding density matrix can then be used to evaluate the expectation of an arbitrary operator.

The outline of the paper is as follows. In Sec.~\ref{S:background}, we introduce an analogy between concepts based on classical probabilities and their quantum analogs, briefly review core concepts, and establish notation.
In Sec.~\ref{S:method} we introduce the  QML and QVI loss function objectives, a strategy to creating a variational family for  density matrices, quantum flows, and detail the optimization procedures~\cite{lattice}. In Sec.~\ref{S:example} we consider the example of a quantum anharmonic oscillator and compare with traditional lattice methodology.  Finally, in Sec.~\ref{S:related} we discuss related work and in Sec.~\ref{sec:disc} we  discuss future directions.

\section{Background and notation}\label{S:background}

\paragraph{Pure states}

Famously, Schr\"odinger's equation describes the behavior of a quantum mechanical particle in terms of a wave function $\psi(x)$ -- a complex valued function of the position $x$ of a particle. While the wave function describes a specific quantum state, it only predicts the probability density of observing the particle at position $x$, which is given by $p(x) = |\psi(x)|^2$.  While it seems natural that one might simply make this identification in translating the classical notions of entropy, free energy, and their ilk to their quantum counterparts, that is not how the story unfolds. 

To see this, it is convenient to think more abstractly and represent a pure quantum state as a vector in a Hilbert space. This vector is essentially just an index over the possible states of the system. Therefore, a pure quantum state is equivalent to assigning unit probability on one index and zero probability to all other indices, which is a quantum system with zero entropy. One might expect that non-zero entropy could be obtained with a quantum mechanical superposition of pure-states; however, that corresponds to vector addition in the Hilbert space and simply results in another pure state with zero entropy.

We will use \emph{bra}-\emph{ket} notation in this paper where a vector living in the Hilbert space is denoted by a \emph{ket}, $\ket{\psi}$, while its hermitian conjugate is denoted by a \emph{bra}, $\bra{\psi}$. The Hilbert scalar product of two vectors $\psi$ and $\phi$ will be denoted by $\langle \psi|\phi\rangle$. In this notation, the wave function of $\psi$ is denoted by $\psi(x) \equiv \langle \psi | x \rangle$, where $| x \rangle$ represent an eigenstate of the position operator with eigenvalue $x$ or, intuitively, a quantum state which is completely localized at position $x$.

\bigskip
\paragraph{Mixed states and the  density matrix}

The extension of classical Shannon entropy to quantum systems, known as the von Neumann entropy, requires a notion of mixed states. Mixed states assign a classical probability to different elements of the Hilbert space. The probabilities assigned to the different states add \textit{incoherently} and behave as classical probabilities.  It is convenient to think of mixed states (in a basis in which the density matrix is diagonal)
as the quantum mechanical analogue of a mixture model, where the mixture coefficients refer to the probabilities associated to the incoherent superposition of pure states, and the pure states correspond to mixture components (see Tab.~\ref{tab:dictionary}).

Mixed states are described by an object known as the density operator or density matrix. Abstractly, the density operator is a positive semidefinite, Hermitian operator of unit trace. The density operator can be written as
\begin{equation}\label{eq:rho}
    \r = \sum_{j} a_j \ket{\psi_j}\bra{\psi_j} \; ,
\end{equation}
where the notation $\proj{\psi_j}$ in Eq.~\ref{eq:rho}, denotes a projector onto the state $\ket{\psi_j}$ and the $a_j \in \mathds{R}^+$ are such that $\Tr[\r] = 1$. If the $\psi_j$ are orthonormal, then to $\sum_j a_j = 1$. 

The density matrix is a concrete representation of this operator in a specific orthonormal basis for the Hilbert space. Often the term density matrix is also used for the abstract operator when the meaning is clear in context. In what follows we will be interested in representing the density matrix in a few specific bases: the (unknown) energy eigenstates $\{\ket{n}\}$, the (known) variational estimates for the energy eigenstates $\{\ket{\tilde{n}}\}$, and the (position) coordinate basis $\{\ket{x}\}$ in which the Hamiltonian is most naturally expressed. In the position basis, the density matrix takes on the concrete form
\begin{equation}\label{eq:rho_xy}
\rho(x,y)  \equiv \bra{x}\rho\ket{y}= \sum_j a_j {\psi^*_j}(x){\psi_j}(y) \; .
\end{equation}

\begin{table}[t]
  \scalebox{0.9}{
  \begin{tabular}{cc}
    \hline\hline
Classical probability                   &       Quantum mechanics  \\
    \hline
    probability density $p(x)$ &     wave function  $\psi(x) \equiv \langle \psi | x\rangle$ \\
    mixture component $p_j(x)$ &     pure state  $\psi_j(x) \equiv \langle \psi_j | x\rangle$ \\
--- & superposition $\ket{\psi} = \sum_j a_j \ket{\psi_j}$ \\
 mixture model $p_\textrm{mix}(x)$   & density matrix $\rho$ \\
\multirow{1}{*}{$p_\textrm{mix}(x) = \sum_j a_j p_j(x)$}  &  $\rho(x,y) = \sum_j a_j {\psi^*_j}(x){\psi_j}(y) $ \\
 $\mathds E[O] = \int_x  p(x) O(x) dx$  &  $\langle O \rangle = \Tr [ \rho \,O ]$  \\
   Gibbs sampling 
   & Monte Carlo approx. of path integral \\
   KL divergence $\KL{p}{q}$  
   & quantum relative entropy $\QRE{\rho}{\sigma}$ \\
    \hline
  \end{tabular}}
  \caption{\small Concepts in classical information theory based on probability densities and their quantum mechanical analogs. 
  }
  \label{tab:dictionary}
\end{table}

\paragraph{The Hamiltonian as a hermitian  operators}\label{sec:ops}

Specific quantum mechanical systems are often specified by the Hamiltonian $\hat{H}$, which is a Hermition operator that corresponds to the total energy of the system and also dictates its time evolution. 
In the examples presented we will consider a simple quantum mechanical Hamiltonian in one dimension
\begin{equation}\label{eq:Hamiltonian}
\hat H = -\frac{1}{2}\frac{d^2}{d{ x}^2}+V({ x}),
\end{equation}
where $V(x)$ is a confining potential, that is $V(x)\to\infty$ for $x\to\pm\infty$.

\bigskip
\paragraph{The thermal density matrix}

Given a Hilbert space of dimension $R$ and a Hamiltonian matrix $\hat{H}$ with energy eigenvalues $\l_n$ and eigenstates $\ket{n}$ such that $\hat{H} \ket{n} = \l_n \ket{n}$, 
the thermal density matrix $\r_T$ associated to $\hat H$ is defined as the trace-normalized exponential of $-\hat H$:
\begin{equation}\label{eq:rho_T}
    \r_T = \frac1Z e^{-\hat  H/T}= \frac1Z \sum_{n=1}^R e^{-\l_n/T} \proj{n},
\end{equation}
where $Z$ is the partition function, a normalizing constant enforcing~$\Tr \r_T = 1$.
In the second equality we have explicitly written the form of $\rho$ in terms of the eigenvalues and eigenstates of $\hat{H}$. The temperature $T$ in this equation determines the relative contribution of the different eigenstates to the thermal density matrix. 
The specific density matrix defined by Eq.~\ref{eq:rho_T} is called the \emph{Gibbs ensemble}.

The exponential factor $e^{-\l_n/T}$ implies that the contribution to the density matrix of eigenstates with large eigenvalues are exponentially suppressed. Thus, at low temperature, an accurate approximation of the density matrix is obtained by keeping only a few eigenstates corresponding to the smallest eigenvalues.

This density matrix describes the equilibrium state of a statistical system at temperature $T$ with Hamiltonian $\hat{H}$~\footnote{Throughout the paper we set the Boltzmann constant $k_B = 1$.}. However, while this physical interpretation of the density matrix can provide some intuition, it is not necessary for the understanding of our methodology. It is clear that if we can learn the thermal density matrix $\r_T$ in the explicit form written in Eq.~\eqref{eq:rho_T}, we can then easily read off the eigenvalues and eigenstates. Therefore, the problem estimating the thermal density matrix in Eq.~\eqref{eq:rho_T} can equally well be thought of as diagonalizing $\hat{H}$. 

\bigskip
\paragraph{The path integral}\label{path_integral}

It is well known that QM can be formulated as a generalization of the stationary action principle of classical mechanics~\cite{landau1}. This is realized via a \emph{path integral} formulation in which the quantum mechanical amplitude for a quantum system to propagate from a state $|{\bf q}_i; 0\rangle$ at time $t=0$ to a state $|{\bf q}_f; \mathcal T\rangle$ at time $t=\mathcal T$ can be written (formally) as a sum over all paths in configuration space such that ${\bf q}(0)={\bf q}_i$ and ${\bf q}(\mathcal T)={\bf q}_f$
\be\label{PI}
\langle {\bf q}_f;T| e^{-i \hat H \mathcal T/\hbar}|{\bf q}_i;0\rangle\sim \int_{{\bf q}(0)={\bf q}_i}^{{\bf q}(\mathcal T)={\bf q}_f} \mathscr D {\bf q}(t) ~e^{i \mathcal{S}[{\bf q},\dot {\bf q}, \ldots]/\hbar}
\ee
where $\hat H$ is the Hamiltonian operator of the quantum mechanical system.
The paths are weighted by an oscillating phase factor given by the exponential of the classical action.
The semiclassical limit of Eq.~{\ref{PI}} is transparent: in the limit $\hbar\to 0$ the path integral is dominated by the saddle point $\delta \mathcal{S}=0$ which then reduces to the stationary action principle of classical mechanics.

Performing the replacement $t\to -i \tau$, a so called Wick rotation, and identifying $\mathcal T\to\beta\equiv \hbar/(k_B T)$, $T$ being  a temperature, the path integral description of the $d$ dimensional dynamical quantum system in Eq.~{\ref{PI}} becomes a statistical description for an associated $d+1$ dimensional system with no time evolution but finite temperature $T$. In particular the Euclidean path integral with periodic boundary conditions calculates the partition function of the quantum system
\begin{align}
Z &=\sum_n e^{-E_n\beta/\hbar }= \sum_n \langle n| e^{-\beta \hat H/\hbar}|n\rangle = \nonumber \\
&= \int_{{\bf q}(0)={\bf q}(\beta)}\mathscr D {\bf q}(\tau) ~e^{-\mathcal{S}_E[{\bf q},\dot {\bf q}, \ldots]/\hbar}
\end{align}
where $\{|n\rangle\}$ is the set of energy eigenstates and 
\begin{equation}
    \label{eq:action}
    \mathcal{S}_E = \int_0^\beta d\tau\, \left[\tfrac12 \rm \dot q(\tau)^2 + V({\rm q}(\tau))\right]
\end{equation}
is the euclidean action.

Given an operator $ O({\bf q})$ representing an observable for the quantum system, the Euclidean path integral also allows to calculate the expectation value of $O$ with respect to the Gibbs ensemble: 
\be\label{gibbsPI}
 \Tr[\rho_T~ O] = \frac{1}{Z}  \int_{{\bf q}(0)={\bf q}(\beta)}\mathscr D {\bf q}(\tau)~  O({\bf q})e^{-\mathcal{S}_E[{\bf q},\dot {\bf q}, \ldots]/\hbar}
\ee

What is subtle in Eq~\ref{gibbsPI} is the meaning to assign to the integration measure $ \mathscr D {\bf q}$, as it requires specifying a measure on a functional space. The formal way to do this is in terms of the Weiner measure~\cite{strocchi}. A more intuitive way to get the same results, and the one we will use in practice in the following, is by discretizing the path ${\bf q}:[0,\beta]\to \mathds{R}^d$ into a finite number of time steps hence turning the infinite dimensional integral into a finite dimensional one. 
Critically, if $\mathcal{S}_E$ has the form in Eq.~\ref{eq:action}, then Eq.~\ref{gibbsPI} can be effectively estimated numerically with Monte Carlo (MC) integration which then provide an empirical estimate of the Gibbs distribution. 
This is done by discretizing the time interval $[0,\beta]$ onto a lattice, and replacing Eq.~\ref{eq:action} by its discrete expression
\be
\hat{\mathcal{S}}_E = \sum_{i=1}^{N_\tau} \left[\frac{({\rm q}(\tau_z)-{\rm q}(\tau_{z-1}))^2}{2 a}+ a\,V\left(\frac{{\rm q}(\tau_z)+{\rm q}(\tau_{z-1})}{2}\right)\right]
\ee
where $\tau_z \equiv \beta\times z/N_\beta$ and $a\equiv \beta/N_\beta$. The infinite dimensional probability space appearing in Eq.~\ref{gibbsPI} has now been replaced by an ordinary finite dimensional one, which can be sampled with traditional MC methods. As a result we can estimate the expectation of any operator $O$ from samples $\{q_i\} \sim \rho_T$ via
\be\label{latticeExp}
 \Tr[\rho_T\,O] \approx  \frac{1}{N_q}  \sum_{i=1}^{N_q}   O({\bf q}_i ) \; .
\ee

\bigskip
\paragraph{An example}

An instructive example is to calculate $\rho_T(y,x)$ explicitly in the simple case of the harmonic oscillator $V(x) = x^2/2$:
\be\nonumber
\rho_T(y,x) =\tfrac{\sinh (\beta/2)}{\sqrt{\pi/2\, \sinh \beta}} \exp\left[-\tfrac{(y^2+x^{2})\coth \beta}{2}+ \tfrac{xy}{\sinh \beta}\right]
\ee
A derivation of this results and an enlightening discussion of the path integral approach to QM can be found in~\cite{rattazziPI}.

Notice that as it could have been guessed, $\rho_T(y,x)$ is a 2 dimensional Gaussian. This property stems from the fact that the action is a quadratic form in the paths and it is lost if anharmonic terms are included in the Hamiltonian. What is less obvious is how this density matrix decomposes uniquely into the an incoherent mixture of energy eigenstates. 

We can rewrite $\rho_T(y,x)$ in terms of energy eigenstates
\begin{align}\label{mixture}
\rho_T(y,x) &= \langle y|\rho_T|x\rangle = \sum_n e^{-E_n/T}  \langle y|n\rangle\langle n|x\rangle \\
&= \sum_n e^{-E_n/T} \psi_n(x)^*\psi_n(y).
\end{align}
The functions $\psi_n(x)$ are the wavefunctions for each of the $n$ energy levels. Such wavefunctions are the Hermite functions 
\be
H_n(x) = \frac{1}{\sqrt{2^n n!\sqrt{\pi}}}e^{-x^2/2} h_n(x),
\ee
where $h_n$ is the $n$th Hermite polynomial
\be
h_n(x) = (-1)^n e^{x^2}\frac{d^n}{dx^n} e^{-x^2}.
\ee
Eq.~\ref{mixture} expresses a very non trivial relation between Hermite functions (which are an orthonormal basis for the Hilbert space of $L^2$ functions on the real line) and the two dimensional Gaussian given by $\rho_T(y,x)$. Unlike a classical mixture model, the density matrix encodes all the details of the quantum states in the ensemble. Thus, estimating the density matrix enables an much richer set of applications than what can be characterized from classical probabilistic summaries of lattice configurations.

\paragraph{Quantum relative entropy.}
The Von Neumann entropy associated to a density matrix $\rho$ is defined as
\be
S=-{\rm{Tr}}[\rho\log\rho]\,.
\ee
In a basis in which $\rho$ is diagonal, $\rho={\rm{diag}}(p_1,p_2, \ldots)$, using the fact that $\log\rho = {\rm{diag}}(\log p_1,\log p_2, \ldots)$ the Von Neumann entropy reduces to
\be
S=-\sum_i p_i \log p_i.
\ee
which is the Shannon entropy of the discrete distribution $p_i$.
A notion of relative entropy can be defined for pair of density matrices. Given two density matrices $\r$ and $\sigma$, one can consider their quantum relative entropy (QRE for short)
\begin{equation}\label{eq:QRE}
    \QRE{\r}{\sigma} \equiv \Tr[\r\,(\log\r-\log\sigma)] \; ,
\end{equation}
which can be thought of as the generalization of the Kullback-Leibler (KL) divergence applied to density matrices~\cite{witten}. The QRE can be thought of as a distance measure between the two density matrices $\r$ and $\sigma$: $\QRE{\r}{\sigma} \geq 0$ and the equality is saturated if and only if $\r = \sigma$. 
In the special case where $\rho$ and $\sigma$ are simultaneously diagonalizable with mixture coefficients given by $p_n$ and $q_n$, their quantum relative entropy reduces to
\begin{equation}\label{eq:entropydiag}
\QRE{\rho}{\sigma} = \sum_n p_n\log \frac{p_n}{q_n}.
\end{equation}
which is exactly the Kullback-Leibler (KL) divergence between the two probability distributions $p$ and $q$.


\section{Method}\label{S:method}

The general idea of our approach to estimating the thermal density matrix relies on minimizing quantum relative entropy between the true, unknown thermal density matrix $\r_T$ and a a member of a variational family of density matrices $\tilde{\r}$. 

We first set up a parametric family of density matrices $\tilde{\r}$ by individually parametrizing each of the eigenstates and eigenvalues of interest
\begin{equation}\label{eq:parametric-rho}
    \tilde{\r} = \frac1{\tilde{Z}} \sum_{n=1}^N e^{-\tilde{\l}_n/T} \proj{\tilde{n}},
\end{equation}
where $\theta$ parametrizes this family of density matrices, $\tilde{Z}$ is an overall normalizing constant to enforce $\Tr \tilde{\r}=1$ and $N<R$ is the number of eigenstates and eigenvalues we are interested in. Throughout this work, any quantity $\tilde{X}$ with a tilde will denote the variational estimate of the quantity $X$. In order for the states $\ket{\tilde{n}}$ to correspond to eigenstates, they need to satisfy an orthonormality condition $\langle \tilde{m}\ket{\tilde{n}} = \delta_{nm}$. The choice of the parametric family of states should be informed by the details of the system. We will discuss this in more detail in the next section. 

Given the parametric family of density matrices $\tilde{\r}$, we can approximate the thermal density matrix $\r_T$ as $\r^*$: the member of the $\tilde{\r}$ family which minimizes the QRE between $\r_T$ and $\tilde{\r}$. Since the quantity $\QRE{\cdot\!}{\!\cdot}$ is not symmetric under the interchange of its two arguments, we have two options: minimizing  $\QRE{\r_T}{\tilde{\r}}$ or minimizing $\QRE{\tilde{\r}}{\r_T}$. In the context of classical probabilities with the standard KL divergence, these two approaches correspond to \emph{Maximum Likelihood } (ML) and \emph{Variational Inference} (VI), respectively. 

The difficulty, of course, is that we don't know $\r_T$ explicitly. Below we will describe how to express the QRE in a way that is still tractable without knowing $\r_T$ explicitly. By using these numerically tractable objectives, the variational optimization proceeds  using standard stochastic gradient descent methods. With $\r^*$ in hand, we can easily read off the eigenstates and the eigenvalues. Since only the ratios $\exp(-\tilde{\lambda}_n/T)/Z$ appear in Eq.~\ref{eq:parametric-rho}, the eigenvalues of $\hat{H}$ can only be extracted from this procedure up to an overall additive constant; however, energy differences can be measured directly.


\subsection{QVI: minimizing $\QRE{\tilde{\r}}{\r_T}$}

Let us first consider optimizing $\QRE{\tilde{\r}}{\r_T}$, which we will call the Quantum Variational Inference (QVI) method. The training objective in this case is:
\begin{align}\label{eq:QVI}
\QRE{\tilde{\r}}{\r_T}&=-\Tr[\tilde{\rho} \log\rho_T]+\Tr[\tilde{\rho} \log\tilde{\rho}]\notag\\
    = \; & \Tr[\tilde{\rho} \hat H] / T  + \Tr[\tilde{\rho} \log\tilde{\rho}] + \log Z,
\end{align}
where in the second equation we have substituted the definition of the thermal density matrix in Eq.~\eqref{eq:rho_T}. The last term, $\log Z$ in this case is independent of $\theta$, and can be dropped from the optimization objective. Further, we notice that the second term is the von Neumann entropy of the thermal density matrix \cite{witten}. Multiplying Eq.~\eqref{eq:QVI} by an overall factor of $T$ we arrive at 
\begin{align}\label{eq:F}
    T \,\QRE{\tilde{\r}}{\r_T} =&\, \Tr[\tilde{\rho} \hat H]  + T\,\Tr[\tilde{\rho} \log\tilde{\rho}] \notag\\
    =&\, \langle \hat H \rangle_{\tilde{\rho}} - T \tilde{S}=\tilde{F}.
\end{align}
When $\hat H$ represents the Hamiltonian of a physical system, $\tilde{F}$ is the free energy associated to the density matrix $\tilde{\r}$. 
Intuitively, this optimization objective is simultaneously minimizing the energy of the thermodynamic system while maximizing the entropy associated to the distribution~$\tilde{\r}$. If all the expression on the right hand side of Eq.~\eqref{eq:F} can be computed numerically, we can directly minimize $\tilde{F}$ using gradient descent in order to estimate eigenvalues and eigenstates $\tilde{\l}$ and $\ket{\tilde{n}}$. The difficulty in this approach is that the term $\Tr[\tilde{\rho} \hat H]$ requires calculating matrix elements for the Hamiltonian in the variational basis states. We will give an explicit example and demonstrate details of this method in Sec~\ref{sec:QVI}. 

\subsection{QML: minimizing $\QRE{\r_T}{\tilde{\r}}$}

The second approach, which we will call the Quantum Maximum Likelihood  (QML) method, is to minimize the alternate form of the quantum relative entropy
\begin{equation}\label{eq:QML}
\QRE{\r_T}{\tilde{\r}}=-\Tr[\rho_T \log\tilde{\rho}]+\Tr[\rho_T \log\rho_T].
\end{equation}
The second term in this equation is independent of the parameter $\theta$ and can be dropped from the optimization objective function. The remaining term $-\Tr[\rho_T \log\tilde{\rho}]$, however, cannot be computed directly as we don't know $\rho_T$ explicitly.

We exploit the analogy for the relationship between the KL distance standard maximum likelihood in which the expectation with respect to an unknown target distribution is approximated with an empirical distribution of samples. In our case we will sample from the thermal density matrix $\rho_T$. Notice however that $\rho_T$ is not a normal probability distribution; however, in the case of quantum mechanical Hamiltonians, sampling from the thermal density matrix is indeed possible using the path integral (see Table~\ref{tab:dictionary}). This makes the QML method a viable alternative to the QVI method in scenarios where sampling is possible but the matrix elements of $H$ cannot be easily computed.

\subsection{Variational density matrices}

A key component of our approach is the construction of a variational family of density matrices $\tilde{\r}$ of the type given in Eq.~\eqref{eq:parametric-rho}, which is expressive enough to capture the $N$ lowest energy states of the system. This comprises of two parts: the variational eigenvalues $\tilde{\l}_n$ and the variational eigenstates $\ket{\tilde{n}}$.

For each state $\ket{\tilde{n}}$ corresponds to an $L^2$ normalized function $\tilde{\psi}_n:\mathds R^d \to \mathds C$. A reasonable parametrization of our variational family, which we pursue in this paper, is obtained by expanding each variational eigenstate $\ket{\tilde{n}}$ onto a set of orthonormal states $\ket{j}$. While the choice of the $\ket{j}$ states would not matter if they formed a complete basis, we are forced to truncate the full infinite dimensional Hilbert space to a finite dimensional subspace. Therefore, a judicious choice of the $\ket{j}$ states will play an important role in the accuracy of the resulting approximation $\tilde{\rho}$.

One can consider $\ket{j}$ to be a subset of the eigenstates of some Hermitian operators $O$, as this ensures picking from an already orthonormal basis. Some of the choices we explore in this papers are the Hermite polynomials (eigenstates of $\hat H$ when $V(q) = q^2/2$) and Fourier modes (eigenstates when $V(q) = 0$ and $q \in [-L,L]$). 

Explicitly we write the parametric family of eigenstates as
\begin{equation}
    \label{eq:state_expansion}
    \ket{\tilde{n}}  = \sum_{j=1}^{M} \tilde{a}_{j,n}\ket j,
\end{equation}
where the $\tilde{a}_{j,n}$ parametrize the variational family and they identify the component of $\ket{\tilde{n}}$ in the $\ket{j}$ direction. Here $M$ represents the number of orthonormal vectors we use to describe our parametric family and coincides with the dimensionality of the subspace spanned by the $\ket{j}$ states. $M$ provides the upper bound on the number of eigenstates we will be able to describe and, at the same time, limits the expressivity of the variational family.

Plugging Eq.~\ref{eq:state_expansion} into Eq.~\eqref{eq:parametric-rho} we obtain the explicit form of out variational ansatz for $\rho_T$:
\begin{equation}
    \label{eq:explicit-rho}
    \tilde{\r} = \sum_{n=1}^N \sum_{j=1}^M \sum_{j'=1}^M \tilde{p}_{n} {\tilde{a}_{j,n}} \tilde{a}_{j',n} \ket{j'}\bra{j}
\end{equation}
where we defined $\tilde{p}_n\equiv \sfrac{1}{\tilde{Z}}\; e^{-\tilde{\lambda}_{n}/T}$ to be the Boltzmann factors.
In order for $\tilde{\r}$ to define a density matrix that can be optimized by minimizing the QRE two conditions have to be met:
\begin{enumerate}
    \item \emph{Orthonormality}: $\bra{\tilde{m}}\tilde{n}\rangle = \delta_{n,m}$.  During optimization, we manually enforce normalization at each step by rescaling the $\tilde{a}_{j,n}$ by the appropriate norm. On the other hand, we impose orthogonality of the states in  Eq.~\ref{eq:state_expansion} by adding a constraint term to the optimization objective 
\begin{equation}
\label{eq:L_perp}
 L_\perp \equiv \sum_{n<m} \bra{\tilde{n}}\tilde{m}\rangle^2 = \sum_{n<m}\Big[\sum_j {\tilde{a}_{j,n}^*} \tilde{a}_{j,m}\Big]^2.
\end{equation}
    
    \item \emph{Unit trace}: $\sum \tilde{p}_n = 1$. This can be implemented by parametrizing the Boltzmann factors by a softmax function.
\end{enumerate}

\subsection{Quantum flows}\label{stateflows}

Given a parametric family of density matrices $\tilde{\r}$ defined as above, we introduce a simple technique that substantially increases the expressivity of the family while respecting the orthonormality and unit trace conditions. This technique is an analogue of normalizing  flows~\cite{larochelle2011neural,papamakarios2017masked,2015arXiv150505770J, DBLP:journals/corr/DinhSB16} on classical densities generalized to an orthonormal basis in a Hilbert space. In other words, while classical flows are constructed to respect $\int p(x) dx =1$, our quantum flows are designed such that the $L^2$ inner product of two states $\langle n\ket m = \int \psi_n^*(x)\psi_m(x) dx$ is preserved. This in particular implies that a set of orthonormal states will flow to another set of states with the same property.

Given an orthonormal basis $\psi_n:\mathds R^d \to \mathds C$ such that $\int \psi_n^*(x)\psi_m(x) dx = \delta_{n,m}$, we define a quantum flow of this basis via the bijection $f^{(\theta)}:\mathds R^d\to\mathds R^d$ as
\begin{equation}
    \label{eq:flow}
 \tilde{\psi}'_n(x) \equiv U[{f^{(\theta)}}; \psi_n]  =  \psi_n(f^{(\theta)}(x)) \det \left| \nabla_x f^{(\theta)}\right|^{\frac12},
\end{equation}
where the factor on the right hand side is the square root of the Jacobian determinant. One can easily check that this ensures that the $L^2$ inner product, orthonormality in particular, is maintained. 

We can use Eq.~\ref{eq:flow} to augment our variational ansatz in Eq.~\ref{eq:state_expansion}. The parameters of this flow-augmented variational family is thus be the union of the parameters used to describe the original eigenstates together with the parameters used to describe the flow function $f^{(\theta)}$. 

The quantum flow augmentation technique has some limitations, however. Eq.~\ref{eq:flow} defines a restricted family of unitary operators. It can be easily verified that the ratio of two different states at the same coordinate cannot be changed via quantum flows alone as the Jacobian determinant cancels.  Also, in one dimension, the number of nodes in a wave function and the total probability of a state in between adjacent nodes is conserved if the bijection is continuous. It is possible to circumvent some of these limitations, for instance the fixed number of nodes of the wave function, by using discontinuous bijections.  Even with these limitations we will see in the experiment section that using quantum flows in conjunction with Eq.~\ref{eq:state_expansion} can lead to a more expressive variational family for a fixed number of basis states and correspondingly more accurate approximations of $\rho_T$.

\subsection{Optimization procedure}
Given the Hermitian operator in Eq.~\eqref{eq:Hamiltonian} and the variational families defined above by Eq.~\ref{eq:explicit-rho} and Eq.~\ref{eq:flow}, we now provide explicit formulae for the two diagonalization methods.

\subsubsection{Quantum Variational inference}\label{sec:QVI}
When minimizing $\QRE{\tilde{\r}}{\r_T}$ we need to compute the two terms in Eq.~\eqref{eq:F}, i.e. the expectation value of $\hat H$ as well as the entropy associated to $\tilde{\r}$. The first depends on both the Boltzmann weights $p_n$ and the coefficients entering the definition of the states in Eq.~\ref{eq:state_expansion}. The latter, if $\tilde{\r}$ is diagonal, is trivially computed as ${S=-\sum p_n \log p_n}$. Strictly speaking, in our case, $\tilde{\r}$ is only diagonal at the minimum of $L_\perp$. However, by initializing the states such that the orthogonality constraint is satisfied and choosing a high enough weight for $L_\perp$ in total loss function, we can ensure that this orthogonality is approximately satisfied during the optimization process.

Including $L_\perp$ from Eq.~\ref{eq:L_perp} the explicit form of the full optimization objective can be written as:
\begin{equation}
    \label{eq:L_VI}
    L_{\text{\tiny QVI}} = \sum_{n,j,j'} \tilde{p}_n {\tilde{a}_{j,n}^*} \tilde{a}_{j',n} H_{jj'} + T \sum_n \tilde{p}_n \log \tilde{p}_n + c_\perp L_\perp,
\end{equation}
where $c_\perp$ is a hyperparameter determining the relative size of the orthogonality constraint. Here, $H_{jj'}=\bra{j}\hat H \ket{j'}$ are the matrix elements of $\hat H$ in the $\ket{j}$ basis. These are independent of the $\theta$ parameters and need only be computed once at the beginning of optimization. Note that if we were to augment Eq.~\ref{eq:state_expansion} by using quantum flows, we would need to reevaluate the matrix element of $\hat H$ on the flow transformed states after each gradient descent update.

\subsubsection{Quantum Maximum Likelihood } 

The alternative form of the QRE, given by Eq.~\eqref{eq:QML}, requires maximizing the quantity $\Tr[\rho_T \log\tilde{\rho}]$.  As mentioned in the previous section, this quantity can be estimated as long as we can treat the thermal density matrix $\rho_T$ which we are trying to learn as an empirical distribution and sample from it. 
As a first step it is useful to rewrite the trace by expanding it on a coordinate basis:
\begin{equation}
    \label{eq:QML_int}
    \Tr[\rho_T \log\tilde{\rho}] = \int dx dy \bra{y}\rho_T\proj{x}  \log\tilde{\rho}\ket{y}.
\end{equation}
Written in this way it is clear that the evaluation of the QRE would be amenable to Monte-Carlo integration if $\rho_T(y, x)\equiv\bra{y}\rho_T\ket{x}$ could be interpreted as a (unnormalized) probability density, from which we could sample.
\begin{eqnarray}
    \label{eq:QML_int3}
    \Tr[&\rho_T& \log\tilde{\rho}] = \int dx dy \bra{y}\rho_T\proj{x}  \log\tilde{\rho}\ket{y} \\ 
      &=& \int dx dy \rho_T(y,x) \sum_{m} \log \tilde{p}_{m} \tilde{\psi}_m(y) \tilde{\psi}_{m}(x) .  \nonumber
\end{eqnarray}

For many Hamiltonian type operators this interpretation exists and it is provided by the path integral formulation of QM~\footnote{It is important to keep in mind that there are many physically relevant example for which such identification is not possible. In particular systems which do not satisfy a time-reversal invariant (as is is the case for instance, for a charged particle in an external magnetic field, or theories with a non vanishing chemical potential), will be such that their euclidean action will not be real and the associated density matrix not readily identifiable with a probability density. A similar problems occurs in lattice simulation and goes under the name of \emph{sign problem}~\cite{Gattringer:2016kco}}. Restricting to the Hamiltonians in Eq.~\ref{eq:Hamiltonian}, we have:
\begin{equation}
    \label{eq:rho_T_est}
    \rho_T(y, x) = \frac{1}{Z} \int_{{\bf q}(0)=x}^{{\bf q}(\beta)=y} \mathscr D {\bf q} ~e^{-\mathcal{S}_E[{\bf q},\dot {\bf q}, \ldots]}\, .
\end{equation}
Here $\beta=1/T$, \mbox{${\bf q}:[0,\beta]\to\mathds \mathscr{X}$} is a path with boundary conditions \mbox{${\bf q}(0)=x$} and \mbox{${\bf q}(\beta)=y$}. 
Eq.~\ref{eq:rho_T_est} can then be understood as a special case of Eq.~\ref{gibbsPI} in which $O = |x; 0\rangle \langle y; \beta|$. 

\begin{figure*}[t]
    \centering
    \includegraphics[clip, trim=0.56cm 0.3cm 0.48cm 0.48cm, width=0.59\textwidth]{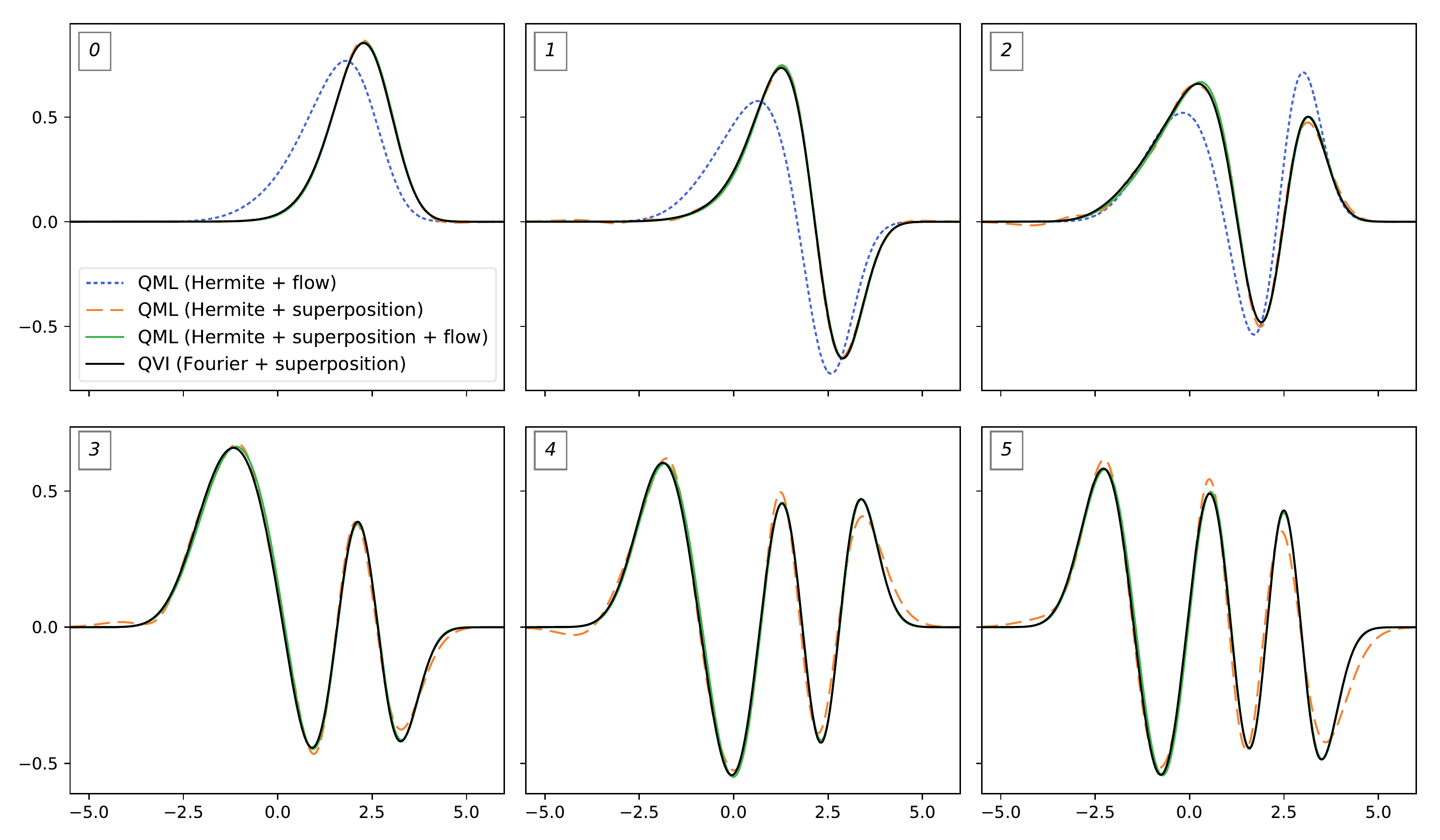}
    \hspace{0.005\textwidth}
    \includegraphics[clip, trim=0.56cm 0.32cm 0.48cm 0.48cm, width=0.387\textwidth]{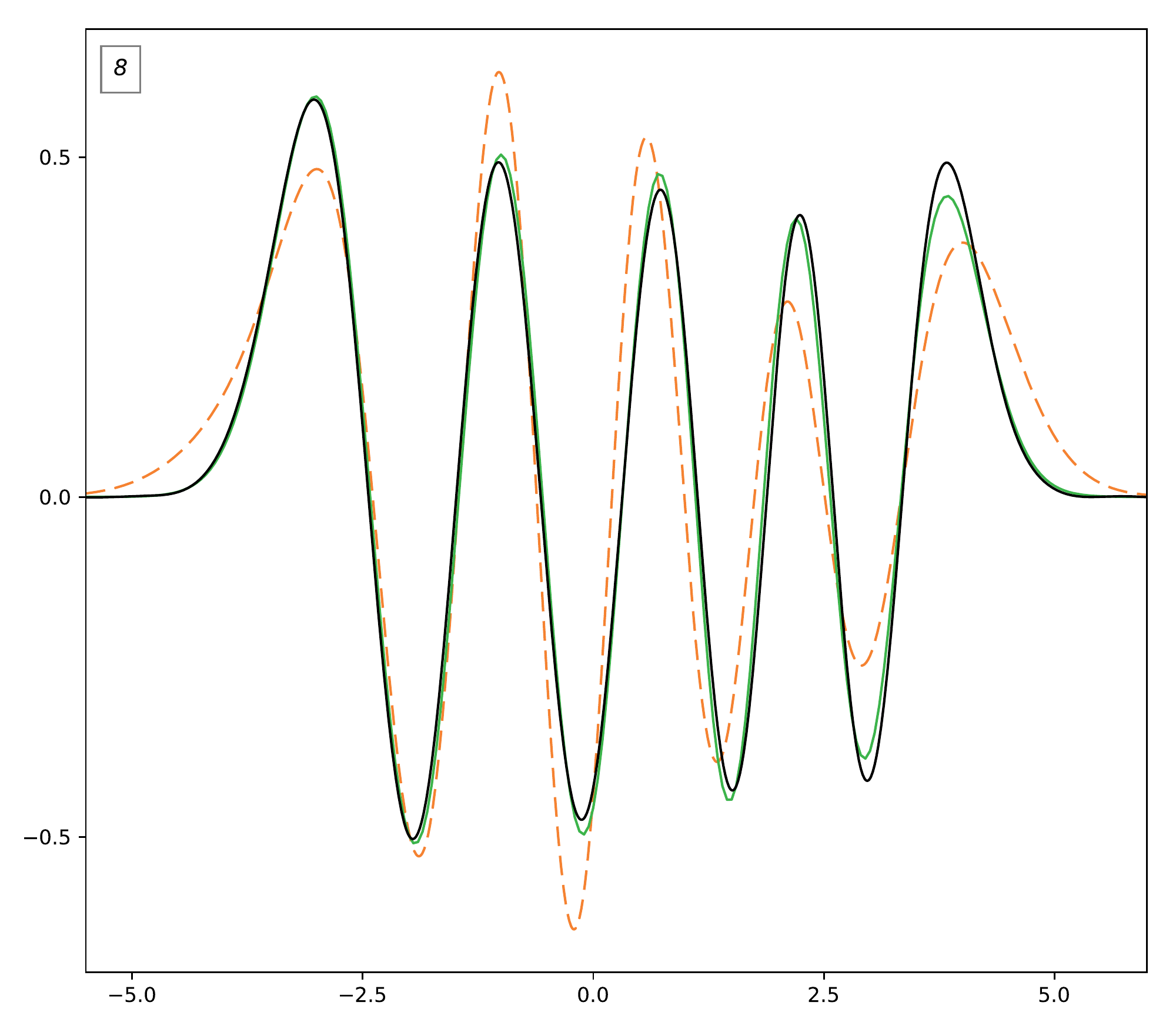}
    \caption{Comparison of estimates of several of the lowest energy eigenstates of the anharmonic oscillator problem estimated using different methods. The results for `QML (Hermite + flow)' (dotted blue lines) are only included in the first 3 eigenstates. The eigenstate number is indicated in the top left of each inset, and we provide a larger format of the $n=8$ eigenstate to reveal the small differences between QML (Hermite + superposition + flow) and the QVI approach. }
    \label{fig:WFs}
\end{figure*}

Given a set of $N_q$ of such paths $\{{\bf q}_i\}_{i=1}^{N_q}$, we can write the empirical approximation to the optimization objective by combining Eqs.~\eqref{eq:rho_T_est} and Eq.~\eqref{eq:explicit-rho}:
\begin{align}
    \label{eq:L_QML}
    L^{\textrm{(emp)}}_{\text{\tiny QML}} = \frac{1}{N_q}\!\sum_{i,n,j,j'} \!\!\log \tilde{p}_n\, {\tilde{a}_{j,n}^*} \tilde{a}_{j',n} {\psi}^*_{j}( y_i ){\psi}_{j'}(x_i) + c_\perp L_\perp,
\end{align}
where $\psi_{j}(x)$ denotes the wave function of the state $\ket{j}$ evaluated at coordinate $x$ and  $x_{i}$ and $y_i$ respectively denote the initial and final end-points of the $i$th sampled path. Notice again that we assume the states describing the variational family to orthogonal in order to evaluate the logarithm of $\tilde{\rho}$ in a closed form. This is approximately enforced by term proportional to $L_\perp$ in Eq.~\ref{eq:L_QML}.

There is an important subtlety here. We are defining the density matrix $\tilde{\rho}$ on the finite dimensional subspace spanned by the $\ket{j}$ states. By inspecting Eq.~\ref{eq:entropydiag}, we see that if the variational density matrix assigns vanishing probability to one of the (unknown) energy eigenstates $\ket{n}$, then the $\QRE{\r_T}{\tilde{\r}}\to \infty$ as $\tilde{p}_n \to 0$. However, if the subspace spanned by $\tilde{\rho}$ does not include $\ket{n}$, then the  term corresponding to $p_n \log \sfrac{p_n}{\tilde{p}_n}$ never appears in the sum and there is nothing stopping the optimization for assigning vanishing probability to the state $\ket{n}$. This is the quantum manifestation of the requirement that  KL divergence $\textrm{KL}[p || q]$ is only defined if $p$ is absolutely continuous with respect to $q$ (i.e. that $q(x)=0$ implies $p(x)=0$).

In order to avoid this problem, we need to extend the support of $\rho_\theta$ to the entirety of the Hilbert space. We do so by assigning a small eigenvalue $\tilde{p}_\perp$ to the whole complement of the subspace spanned by the $\ket{j}$ states, which will act as a regularizer:
\begin{equation}\label{eq:rho_comp}
    \tilde{\rho} = \sum_{n=1}^ N \tilde{p}_n \proj{\tilde{n}} + \tilde{p}_\perp\Big[\mathds{1}  - \sum_{n=1}^ N \proj{\tilde{n}} \Big]\,.
\end{equation}
For consistency, $\tilde{p}_\perp$ needs to be smaller than all the $\tilde{p}_n$ we are including in the expansion of the truncated density matrix. Because the two terms in Eq.~\eqref{eq:rho_comp} are orthogonal, we can compute the logarithm as
\begin{equation}\label{eq:log_rho}
    \log\tilde{\rho} = \mathds{1} \log \tilde{p}_\perp + \sum_{n=1}^ N \log\frac{\tilde{p}_n}{\tilde{p}_\perp}\, \proj{\tilde{n}}.
\end{equation}
In practice, the result of this correction is to disfavor exactly the situation described above in which the variational family collapses in the complement of the states spanned by the main eigenstates. We also notice that this adjustment is not needed in the QVI method as the absolute continuity condition is satisfied in $\QRE{\tilde{\r}}{\r_T}$.

The empirical loss Eq.~\eqref{eq:L_QML} can be minimized by gradient simple gradient descent. We use Markov Chain Monte Carlo (MCMC) to sample the paths according to the measure defined by Eq.~\ref{eq:rho_T_est}. In particular we use affine MCMC~\cite{goodman2010} and its implementation in~\cite{mcmchammer}.

\section{Example: Anharmonic oscillator}\label{S:example}

We demonstrate our methodology by estimating the eigenstates corresponding to the ten smallest eigenvalues of a Hamiltonian of the form Eq.~\eqref{eq:Hamiltonian} with potential function $V(x)$ given by \mbox{$V(x)=x^4/16-x^2/2-x$}. 

For the QML method, we use three different variational families. First, we expand the parametric family of states in terms of Hermite (see Sec.~\ref{path_integral} for their definition) functions $H_j(x)$ up to $j=10$: $\tilde{\psi}_n(x) = \sum_{j}\tilde{a}_{j,n} H_o(x)$, second we augment this expansion using a parametric quantum flow: $\tilde{\psi}_n(x) = \sum_{j<10}\tilde{a}_{j,n} U[f^{(\theta)},H_j]$, and third we only use flows on the Hermite functions without allowing any mixing of the states, i.e. $\tilde{\psi}_n(x) = U[f^{(\theta)},H_n]$. We use stochastic gradient descent with $500$ paths in each gradient step and $2\times 10^5$ optimization steps, equaling a total of $10^8$ individual paths sampled. For the orthogonality constraint coefficient we use $c_\perp=10^2$. The explicit form of the quantum flow we use is detailed in Appendix~\ref{flows}. 

We notice that the size of the batches is critical for convergence of the QML method. If batch size is lowered (halved in this case) the variance of the empirical estimate is too large and we find no convergence. This can be traced back to the fact that the Trace used for the expectation with respect to $\rho_T$ in Eqns.~\ref{eq:QML_int} and~\ref{eq:QML_int} are approximated with a sum over samples from the empirical distribution. This sum serves two purposes: the first is to estimate the expectation of the argument $\log \tilde{p}_{m}$ and the second is to implement the projection of $\rho_T$ onto the corresponding state $\ket{\tilde{m}}$. Sufficient samples are needed so that the projection $\langle n \ket{\tilde{m}}$ implemented as a Monte Carlo integral in the $x$-domain is sufficiently accurate.

For the QVI method, we expand the states in terms of the first 40 Fourier modes on the interval $[-L, L] \equiv [-10,10]$ with $\psi_{j=0}(x) = \sfrac{1}{\sqrt{2 L}}$ and
\begin{equation}
\psi_{2j-1}(x) = \frac{1}{\sqrt{L}}\sin\left(\frac{j \pi x}{L}\right) ,~\psi_{2j}(x) = \frac{1}{\sqrt{L}}\cos\left(\frac{j \pi x}{L}\right)
\end{equation}
for $j=1,\ldots, 20$.
For the orthogonality constraint coefficient we let $c_\perp=10^3$. In both methods, we use Adam optimizer with learning rate $10^{-3}$. 

\begin{figure}[ht]
    \centering
    \includegraphics[clip, trim=0.25cm 0.25cm 0.25cm 0.2cm, width=0.47\textwidth]{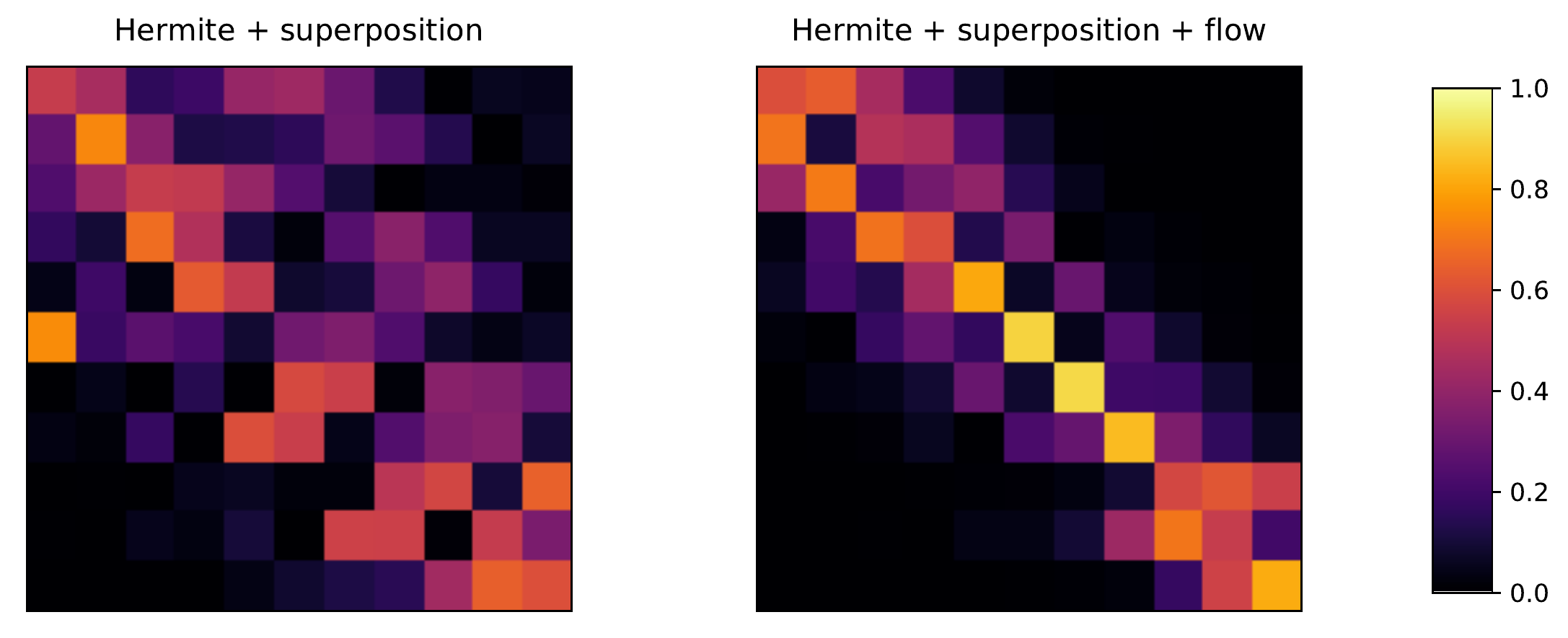}
    \caption{The  magnitude of the expansion coefficients $\tilde{a}_{j,n}$  of the putative eigenstates expanded in terms of the Hermite functions before (left) and after (right) implementing quantum flows.}
    \label{fig:spread}
\end{figure}

The results for the eigenvalue and eigenstate computations are given in 
Fig.~\ref{fig:WFs}. The results of the QVI method are indistinguishable from brute force diagonalization results (which are feasible in this simple example, but which do not scale to larger systems) up to our working precision of $10^{-5}$, both in terms of eigenvalues and eigenstates.  The QML method with quantum flows applied to the Hermite polynomials for the corresponding energy eigenstate of the simple harmonic oscillator performs the worst. This is expected due the limitations associated to flows that we discussed in Sec.~\ref{stateflows}. On the other hand, once we allow for superpositions of the quantum flows we see a dramatic improvement in the accuracy of the variational family. This improvement is noticeable as early as the first excited state where we can see wobbles in the tails of the mixed Hermite results. 

 Augmenting the expressivity of the parametrized family with flows can dramatically increase the accuracy of the eigenvalue estimation, especially for the higher excited states. Fig.~\ref{fig:spread} shows the magnitude of the expansion coefficients $\tilde{a}_{j,n}$ with and without quantum flows. We can see that using quantum flows, results in a more efficient approximation of the eigenstates. Instead of modeling the perturbations of the anharmonic potential as a complicated superposition of eigenstates for the simple harmonic oscillator, the quantum flows are able to perturb the basis states themselves. This is manifest  with the $\tilde{a}_{j,n}$ for the right plot being more closely to be proportional to the diagonal $\delta_{j,n}$. This illustrates how quantum flows applied to a basis known to be relevant to the system provide an opportunity to inject expert knowledge while maintaining expressivity in the variational family. The relative error between the estimated eigenvalues and eigenvectors of the QML and QVI methods is given in Fig.~\ref{fig:EVs}.

\subsection{Comparison to traditional lattice approach} 

Numerical lattice techniques~\cite{lattice} (see also \cite{Lepage:1998dt} for a very pedagogic introduction) can also be used to approximate the energy levels of the Hamiltonian in Eq.~\ref{eq:Hamiltonian}. 

The starting point is the evaluation of the $\tau$ dependence of the correlation function of judiciously chosen operators $O$
\begin{equation}\label{corrlattice}
C(\tau;T)\equiv \langle O(0)O(\tau)\rangle_{\rho_T}
\end{equation}
where $0\leq\tau\leq \beta\equiv 1/T$. These correlation functions play the role of summary statistics and the choice of operators is analogous to feature engineering in machine learning applications. 

Lattice technique use the fact that Eq.~\ref{corrlattice} can be written in two different ways. The first one is in terms of the Gibbs ensemble density matrix $\rho_T$ as
\be\label{tracelattice}
C(\tau;T) = \Tr [\rho_T\, O(0)O(\tau)].
\ee
By expanding $\rho_T$ on an energy basis and using time evolution to write $O(\tau) = e^{\hat H \tau} O e^{-\hat H \tau}$, $O\equiv O(0)$, the trace in Eq.~\ref{tracelattice} can be rewritten as
\be\label{rholattice}
C(\tau;T) =  Z^{-1}\sum_n e^{-E_n\beta}\sum_m e^{-(E_m-E_n)\tau}|\langle n |O|m\rangle|^2,
\ee
exposing the contribution of the various eigenstates to the correlation function. A second way to write  Eq.~\ref{corrlattice} (and the way lattice actually evaluates it) is through a path integral. Similarly to Eq.~\ref{gibbsPI} the path integral representation of Eq.~\ref{corrlattice} is given by
\be\label{pilattice}
 C(\tau;T)= Z^{-1}\int dx \int_{{\bf q}(0)=x}^{ {\bf q}(\beta) =x} \mathscr D {\bf q} ~O(0)O(\tau)~e^{-\mathcal{S}_E[{\bf q},\dot {\bf q}, \ldots]}.
\ee
Notice the integral only extends over periodic paths as a consequence of the cyclic nature of the trace in Eq.~\ref{tracelattice}. When $O$ can be expressed as a function of the path integrals variables, {\bf q} in our case, Eq.~\ref{pilattice} can be evaluated numerically using similar techniques to those we used for QML.

In order to extract the energy of the first excited state of the anharmonic oscillator hamiltonian we set $O={\rm{q}}$. We put Eq.~\ref{rholattice} in a more symmetric form by considering $\bar \tau \equiv \tau - \beta/2$. Eq.~\ref{rholattice} becomes
\be\label{taubar}
 C(\tau;T)= Z^{-1} \sum_{n,m} e^{-(E_n+E_m)\beta/2}e^{-(E_m-E_n)\bar \tau}|\langle n |{\rm{q}}|m\rangle|^2
\ee
For large $\beta$ (or equivalently small temperature $T$), the leading contributions to~Eq.~\ref{taubar} comes from the ground state and the first excited state, all other terms being further exponentially suppressed
\begin{align}\label{tracelattice2}\nonumber
 C(\tau;T)\approx & ~ 2~ Z^{-1}e^{-(E_1+E_0)\beta/2}\cosh(\Delta E\,\bar\tau)|\langle 1 |{\rm{q}}|0\rangle|^2 \\&+ Z^{-1}e^{-E_0\beta}|\langle 0 |{\rm{q}}|0\rangle|^2
\end{align}
In order to extract $\Delta E=E_1-E_0$ we then parametrize Eq.~\ref{pilattice} as 
\be\label{fitform}
C(\tau;T) = A \cosh[\Delta E\, (\tau-\beta/2)]+B
\ee
and fit $A$, $B$ and $\Delta E$ to the lattice data. We do this in practice by fixing $\beta=1/T=10$ and evaluating $C(\tau;T)$ by sampling paths from the euclidean action. We discretize the $[0,\beta]$ time interval on lattice of 160 equally spaced points. By performing a $\chi^2$ fit to Eq.~\ref{fitform} we obtain $\Delta E = 1.58\pm 0.01$.

This lattice estimate was \emph{not} made using the same paths that were sampled for the QML method. As already explained, the paths required for lattice simulations and QML satisfy different boundary conditions in the time direction. The different boundary condition corresponds to a factor of two in temperature, which complicates a fair comparison of the computational costs of the methods; however, we use roughly equal computing resources for both QML and lattice methods. 

\begin{figure}[ht]
    \centering
    \includegraphics[clip, trim=0.2cm 0.2cm 0.3cm 0.2cm,width=0.45\textwidth]{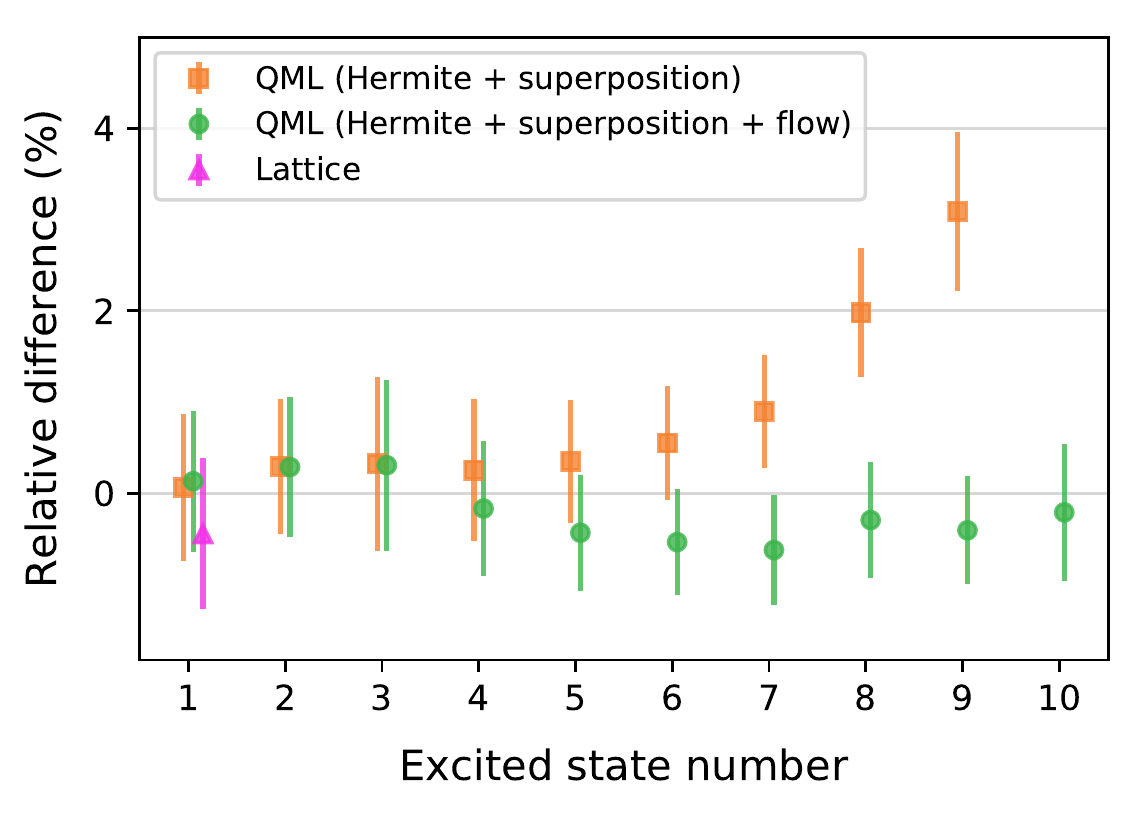}\\
    \includegraphics[clip, trim=0.2cm 0.2cm 0.3cm 0.2cm,width=0.45\textwidth]{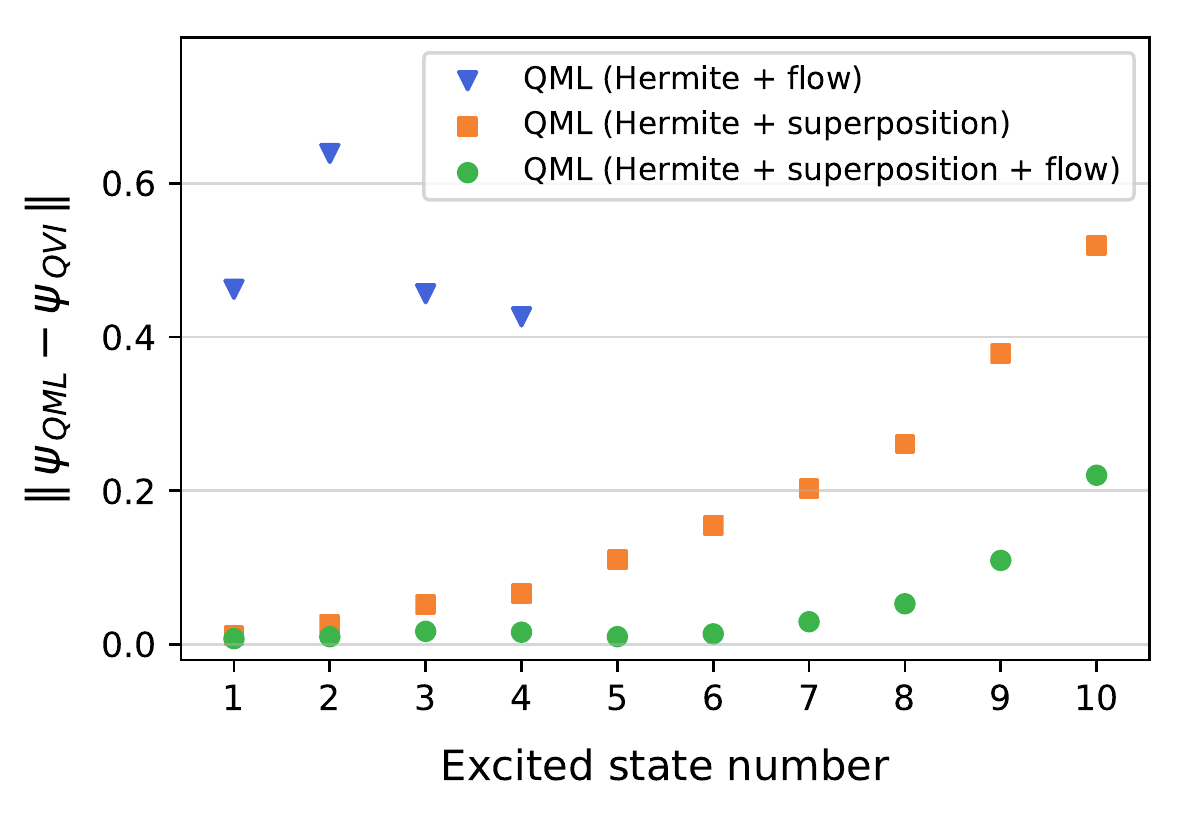}
    \caption{\emph{Top panel}: fractional difference between QML and QVI for the eigenvalues of the first 10 energy for the anharmonic oscillator Hamiltonian. \emph{Lower panel}: $L^2$ distance between QML and QVI eigenstates for the anharmonic oscillator Hamiltonian, $||\psi||^2\equiv \int\, dx|\psi(x)|^2$.}
    \label{fig:EVs}
\end{figure}

As can be seen from Fig.~\ref{fig:EVs}, the estimate and accuracy of these sampling based methods for the first excited state are comparable. Extraction of higher energy levels and operators matrix elements is also possible by fitting multiple correlators simultaneously, and various methods to do this have been developed by the lattice community~\cite{Lin:2007iq}.
However, at least for the restricted class of hamiltonians that we are considering in this paper, we view the ability to extract eigenstates and eigenvalues simultaneously by optimizing a single objective function as an advantage of the QML and QVI methods compared to traditional lattice techniques. In particular, the resulting estimate of the density matrix $\tilde{\rho}$ can then be used to evaluate the expectation $Tr[\tilde{\rho} O]$ of an arbitrary operator. 

From the point of comparing systematic uncertainties of the approaches described in this paper and the lattice, they both share the necessity of an infrared and ultraviolet cutoff. For the example presented in this paper these regulators arise from the discretization of the time direction. Although improvable, such approximation limit the accuracy of the observable which are extracted by both methods. The nature of the fit is quite different in the two cases as traditional lattice techniques fit the $\tau$-dependence of the correlation function $C(\tau; T)$, while QML and QVI work natively in the space of the lattice configurations $\mathbf{q}$. In the limit where the variational family has infinite capacity, QML and QVI provide a path towards asymptotically exact solutions. In realistic finite-sample and finite-capacity situations there will be a tradeoff between bias and variance. The limited capacity of the variational family may introduce some bias or systematic uncertainty, but the total uncertainty budget may be reduced due to improved sample efficiency.

\section{Related work}\label{S:related}

\paragraph{Ground states of quantum systems.} 
In the context of using machine learning for quantum many body systems, there has been recent activity in using machine learning techniques to create variational families for finding the ground state energy for many body quantum systems (eg. Ref.~\citep{carleo_solving_2017}). In this case, the goal is usually to minimize the ground state energy $\sfrac{\langle\tilde{\psi}_0 | H | \tilde{\psi}_0\rangle }{\langle\tilde{\psi}_0 | \tilde{\psi}_0\rangle}$. Restricted Boltzman machines have been used extensively for parametrizing the complex coefficients associated to basis states of the discrete Hilbert space, which is grows exponentially with the number of particles in the system. Our work is similar in spirit, but we model an entire density matrix and our goal is to approximate the thermal density matrix for the system. 

\paragraph{Eigen-decomposition.} 
Our work was motivated by solving quantum systems, but can be reformulated generically in terms of estimating eigenvalues and eigenfunctions of linear operators. A wealth of literature exists for addressing this classic problem. For discrete systems of size $n$, a full eigen-decomposition can be achieved in  $\mathcal{O} (n^3)$~\cite{BruteForce}. However, as $n$ grows the scaling of these methods quickly make them untenable. In cases where $n$ is so large that the matrix itself does not fit into memory, iterative  techniques are generally used which  can efficiently compute a fixed number of eigenvectors by repeated applications of matrix-vector products~\cite{Iterative-Diag}. However, when the eigenstates  are continuous (e.g.\ eigenfunctions in continuous Hilbert spaces) or they are expressed in an exponentially large basis (e.g.\ spin states on large lattices), it becomes impossible to even express the entirety of a single eigenstate in memory. To address this case, approximation tools have been introduced, which recreate the entirety of the eigenstate from a finite sampling via interpolation~\cite{Nystrom-Diag}. These kernel-based interpolation schemes, however, do not take advantage of an expert's prior knowledge regarding the general form of the eigenstates. Shortly after an initial version of this work was presented in Ref.~\cite{slides}, a related technique known as Spectral Inference Networks (SPIN) was proposed, which was motivated by the general eigen-decomposition problem~\cite{GBrain-diag}. That work also describes an iterative procedure and they apply it to the Hamiltonian of a 2-D hydrogen atom as well as non-quantum systems such as video in the context of slow-feature analysis. The authors of that work comment that constructing an explicitly orthonormal function basis (as in the case of our quantum flows) may not be possible in the general setting where the base measure needed to evaluate the inner product is not known; however, in the quantum mechanical setting we consider this is not a barrier.

\section{Discussion}\label{sec:disc}
Some other venues of future exploration currently under consideration are as follows:

\paragraph{Quantum flows.} 
In this work we introduced quantum flows, an extension of the normalizing flows on classical probability densities to orthonormal states. While the unitary operators described by the change of variables are a quite restricted and have some limitations, they still can increase the expressivity of the variational family. It is interesting to consider the continuous-time limit of composing multiple bijections where the Jacobian trace can be calculated if the transformation is specified by an ordinary differential equation. This approach has been recently studied for classical probability densities in the FFJORD algorithm~\cite{Grathwohl2018FFJORDFC}. Similarly, in many quantum systems with translational symmetry, convolutional architectures are natural and provide a powerful inductive bias on the form of the variational family. Recently, invertible (bijective) convolutions have been developed as a new class of normalizing flows that admit an exact likelihood (density)~\cite{kingma2018glow,DBLP:journals/corr/abs-1901-11137}. 
Finally, invertible ResNets have also been developed, which extend the use of this powerful variational family to problems that require a tractable likelihood~\cite{DBLP:journals/corr/abs-1811-00995}. It should be straightforward to utilize these recently developed classical flows in the quantum setting via Eq.~\ref{eq:flow}.

More generally, it would be interesting to extend the notion of quantum flows to the rich literature on quantum circuits~\cite{908999} and connect with approaches to learning approximation of unitary matrices factorized in terms of Givens rotations~\cite{DBLP:journals/corr/MathieuL14}.

\paragraph{Other distance metrics.} While both the QML and QVI approaches are based on the quantum relative entropy -- the analog of the KL divergence. When the target density and the  density of the variational model are very different, the KL distance can be unstable as it involves a ratio of the densities. Moreover, there is little gradient signal in these situations.  Recent work in generative modeling from the machine learning community has explored alternative distance metrics on the space of probability distributions. In particular,  optimal transport or Wasserstein distances have been explored as they alleviate many of the issues with the KL divergence. Quantum analogues of the Wassterstein distance exist, and it would be interesting to compare their performance in this context~\cite{carlen2017gradient,Chen2017WassersteinGO,chen2018matrix}. Furthermore we note the connection between the dynamical systems point of view in FFJORD~\cite{Grathwohl2018FFJORDFC} and the flows on density matrices described in Ref.~\cite{chen2018matrix}.

\paragraph{Extension to field theory.} This work was originally motivated by considering the traditional approaches used in lattice quantum chromodynamics (LQCD) in the light of contemporary approaches to inference with implicit models that do not admit a tractable density or posterior. Current methods in LQCD are completely non-parametric and are based on ensemble averages of operators as in Eq.~\ref{corrlattice}. In the case of LQCD, one is primarily interested in probing energies and matrix elements of current operators in a few low-lying states, and fluctuations in the lowest energy states lead to high variance estimators for quantities associated to higher energy states. Therefore, extensions of this approach from the quantum mechanical setting described here to quantum filed theory is a worthwhile subject of future study. 
 
\paragraph{Application to Hamiltonian renormalization.} The practice of truncating the Hamiltonian of a quantum mechanical system to a finite dimensional subspace, while adjusting the parameters to keep the low energy physics unaltered,  goes by the name of Hamiltonian truncation (or Truncated Spectrum Approach). It was popularized in the 90's after the works of Yurov and Zamolodchikov~\cite{TH1,TH2} and has had great success in a number of different applications (for a recent review see~\cite{TH-review}). Similarly, Density Matrix Renormalization~\cite{white1992density} and Entanglement Renormalization~\cite{vidal2007entanglement} involve a similar truncation of the Hilbert space. 
It would be of great interest to see if a variational truncation of the Hamiltonian as described in this work can improve upon the numerical accuracy of these techniques.

\paragraph{Conclusion.}
In this paper, we introduce QML and QVI,  two methods for estimating the density matrix for a quantum system. 
The approach is qualitatively different than traditional lattice techniques that rely on the time dependence of correlation functions that summarize the lattice configurations. In contrast, QML and QVI work natively in the space of the lattice configurations $\mathbf{q}$ and allow for the extraction of eigenstates in addition to eigenvalues. The resulting estimate of the density matrix can then be used to evaluate the expectation of an arbitrary operator, which we view as an advantage of the QML and QVI methods compared to traditional lattice techniques.

\vspace{1in}

\begin{acknowledgments}
We thank Aida  El-Khadra, Enrico Rinaldi, Michael Albergo, Jaan Altosaar, Rajesh Ranganath, Johann Brehmer, Massimo Porrati, Dries Sels, Giuseppe Carleo, Nima Arkani-Hamed, and Juan Maldacena for insightful discussions. We thank Phiala Shanahan for comments on the manuscript. KC would like to thank the Institute for Advanced Study for providing such a conducive environment for research on this topic and the Moore Sloan Data Science Environment for fostering this type of interdisciplinary research. KC and DP were supported through the NSF grant ACI-1450310.  SG is supported by the James Arthur post-doctoral fellowship. 
\end{acknowledgments}

\bigskip

\appendix

\section{quantum flows}\label{flows}
The specific implementation of the one-dimensional quantum flows employed in the QML approach presented here uses a very simple form for the bijection $f$. This simple form provides two necessary desired properties, expressivity and monotonicity. Given an interval $[a,b]$ we consider the sublattice $\{x_i | x_i \equiv a+(b-a)\times i/n, 0\leq i\leq n\}$. We then construct
\begin{equation}
f^{(C)}(x)\equiv \sum_{i=0}^n C_i\tanh(x-x_i).
\end{equation}
with the parameters $\theta$ of the flow being the non-negative constant $C_i$. This form results in extremely simple gradients
\begin{equation}
\frac{\partial f^{(C)}(x)}{\partial C_i} = \tanh(x-x_i).
\end{equation}
In the anharmonic oscillator experiment, we use $n=400$ points in the interval of $[-10,10]$, giving a sublattice spacing of $0.05$.

\bibliographystyle{apsrev4-1}
\bibliography{biblio}
\end{document}